\documentclass[preprint,aps,superscriptaddress,nofootinbib,tightenlines]{revtex4}

\usepackage{epsfig}

\def\OMIT#1{{}}

\def\cf{{\cal F}}

\newcommand{\gsim}{\raisebox{-0.7ex}{$\stackrel{\textstyle >}{\sim}$ }}
\newcommand{\lsim}{\raisebox{-0.7ex}{$\stackrel{\textstyle <}{\sim}$ }}

\begin{document}

\preprint{\vbox{
\psfig{file=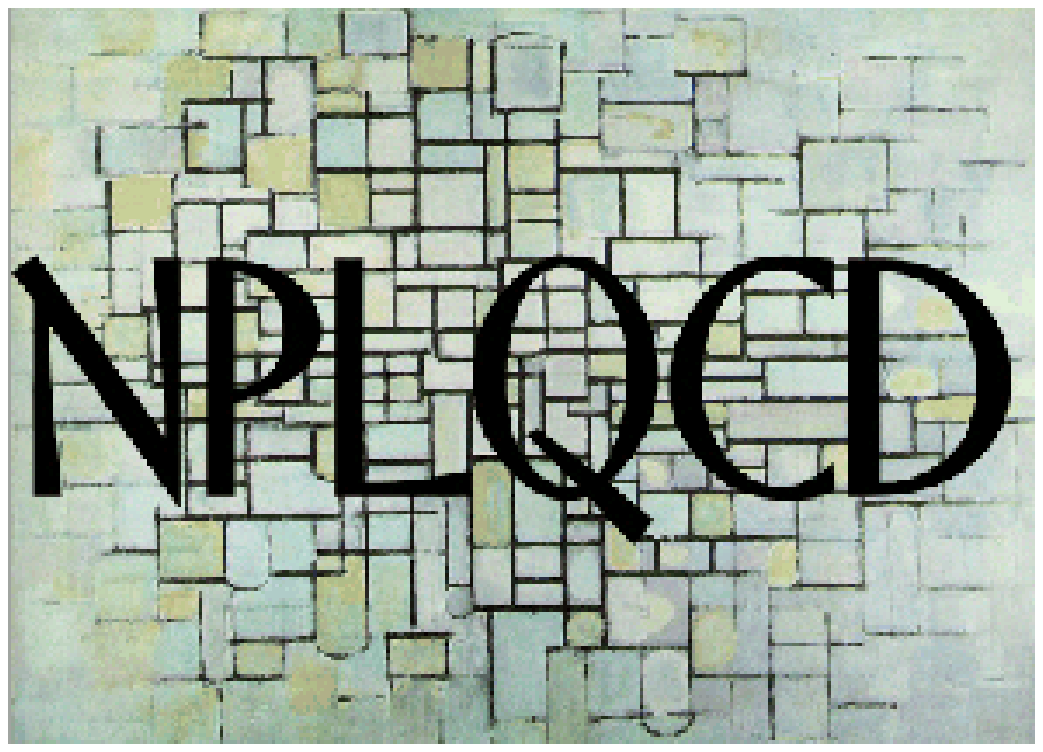,width=0.8in,angle=0}\hfill
\hbox{UNH-04-01}}}
 
\phantom{ijk}
\vskip 0.5cm
\title{Nucleon Masses and Magnetic Moments in a Finite Volume}

\author{\bf Silas R.~Beane}
\affiliation{Department of Physics, University of New Hampshire,
Durham, NH 03824-3568.}
\affiliation{Jefferson Laboratory, 12000 Jefferson Avenue, 
Newport News, VA 23606.}

\vphantom{}
\vskip 0.5cm
\begin{abstract} 
\noindent We compute finite-size corrections to nucleon masses and
magnetic moments in a periodic, spatial box of size $L$, both in QCD
and in partially-quenched QCD. 
\end{abstract}

\maketitle

\section{Introduction}
\label{sec:intro}

\noindent Impressive progress is being achieved in deriving properties
and interactions of hadrons using lattice QCD. In several instances,
lattice methods are making predictions of hadronic quantities at the
several-percent level~\cite{Lepage}. Despite remarkable technical advances,
current computational limitations continue to 
necessitate the use of quark masses, $m_q$, that are significantly
larger than the physical values, lattice spacings, $a$, that are not
significantly smaller than the physical scales of interest, and lattice
sizes, $L$, that are not significantly larger than the pion Compton
wavelength~\cite{Jansen}. Therefore, lattice QCD simulations of hadronic physics
require extrapolations in the quark masses, lattice spacing and
lattice size, and ultimately it is confidence in these extrapolations
that will allow a confrontation between lattice QCD and
experiment. Fortunately, in many cases, the dependence of hadronic
physics on these parameters can be calculated analytically in the
low-energy effective field theory (EFT).  Calculability requires maintaining the
hierarchy of mass scales,
\begin{equation}
|{\vec p}\;|\; ,\; m_\pi\ \ll \  \Lambda_\chi\  \ll \ a^{-1}\ \ ,
\label{eq:hierarchy}
\end{equation}
where $|{\vec p}\;|$ is a typical momentum in the system of interest,
$m_\pi$ is the pion mass and $\Lambda_\chi\sim 2\sqrt{2}\pi f$ is the
scale of chiral symmetry breaking ($f=132~{\rm MeV}$ is the pion
decay constant).  In a spatial box of size $L$, momenta are quantized
such that ${\vec p}=2\pi{\vec n}/L$ with ${\vec n}\in{\bf Z}$. The
hierarchy of eq.~(\ref{eq:hierarchy}) then requires maintenance of the
additional inequality $f L\gg 1$. This bound ensures that (non-pionic) hadronic physics
lives inside the box. In addition, the bound $(m_\pi L)^2(f L)^2\gg 1$
ensures that the box size has no effect on spontaneous chiral symmetry breaking~\cite{Leutwyler:1987ak,Leutwyler:1992yt}. 
These two bounds, taken together, then imply that we must have $m_\pi L\gsim 1$. 
When $(m_\pi L)^2(f L)^2\lsim 1$, and therefore $m_\pi L\ll 1$, momentum zero-modes must be treated 
nonperturbatively~\cite{Leutwyler:1987ak,Leutwyler:1992yt} and one is in the so-called
$\epsilon$-regime. 

Here we will consider the range of pion masses, $139~{\rm
MeV}<m_\pi<300~{\rm MeV}$, and therefore we will take $L\gsim 2~{\rm
fm}$, keeping in mind that the EFT may be reaching the limits of its
validity when this bound on $L$ is saturated, particularly when the
pions are light. For the observables considered here, finite-volume
effects tend to be small for $L>4~{\rm fm}$. It is therefore of
interest to have control over the finite-size dependence of hadronic
observables in the range $2~{\rm fm}<L\leq 4~{\rm fm}$.  Chiral
perturbation theory ($\chi$PT), which provides a systematic
description of low-energy QCD near the chiral limit, is the
appropriate EFT to exploit the hierarchy of eq.~(\ref{eq:hierarchy})
and to describe the dependence of hadronic observables on
$L$~\cite{Leutwyler:1987ak,Gasser:1987zq,betterways,Luscher}.  Recent
work has investigated the finite-volume dependence in the
meson~\cite{Colangelo:2002hy,Colangelo:2003hf,S92,golter1,Pqqcd2,golter3,Leinweber:2001ac,davidlin,Becirevic:2003wk,ArLi}
sector and in the
baryon~\cite{AliKhan:2002hz,AliKhan:2003kb,AliKhan:2003rw,Khan:2003cu,Kronfeld:2002pi}
sector.  In this paper we compute the leading finite-volume dependence
of the nucleon masses and magnetic moments in baryon $\chi$PT,
including the $\Delta$ as an explicit degree of freedom~\footnote{
Recent work~\cite{hemmweise,hemmver} has suggested that for certain observables,
a rearrangement of the chiral expansion may improve convergence. We do not utilize
these modified chiral expansions in this paper.}. The
finite-size dependence of the nucleon mass was first studied in
Ref.~\cite{Luscher}, and has recently been computed to $O(m_\pi^4)$
in baryon $\chi$PT (without including the $\Delta$ as an explicit degree
of freedom) in Ref.~\cite{Khan:2003cu}. (Some discussion of the effects
of the $\Delta$ on the finite-size dependence of the nucleon mass
appears in Ref.~\cite{AliKhan:2002hz}.)

We also give expressions for the finite-size dependence of the nucleon
masses and magnetic moments in partially-quenched QCD (PQQCD),
including strong isospin breaking.  The cost of simulating dynamical
quarks with light masses suggests separately varying the sea and
valence quark masses in the lattice QCD partition function, a
procedure known as partial-quenching. This procedure has important
advantages beyond issues of cost; by increasing the dimensionality of
the parameter space that is explored, lattice QCD simulations can
provide additional ``data'', which can significantly improve the
quality of extrapolations. $\chi$PT has been extended to describe both
quenched QCD (QQCD)~\cite{Sharpe90,S92,BG92,LS96,S01a} with quenched
chiral perturbation theory (Q$\chi$PT) and
PQQCD~\cite{Pqqcd1,Pqqcd2,Pqqcd3,Pqqcd4,SS01} with partially-quenched
chiral perturbation theory (PQ$\chi$PT).  Recently, meson and baryon
properties have been studied extensively in both
Q$\chi$PT~\cite{LS96,S01a} and
PQ$\chi$PT~\cite{CSn,BSn,BSpv,Leinweber:2002qb,Arndt:2003ww}.
The effective field theory (EFT) describing the low-energy dynamics of
two-nucleon systems and nucleon-hyperon systems in PQQCD has also been
explored~\cite{BSnn,ABSl,Beane:2003yx}.

This paper is organized as follows. In Section~\ref{sec:masses}, the
leading finite-size corrections to the nucleon masses are
computed. The same is done for the nucleon magnetic moments in
Section~\ref{sec:magmoms}.  We conclude in
Section~\ref{sec:conc}. Mathematical details and the
partially-quenched extensions of the QCD results (including strong
isospin breaking) are left to Appendices.

\section{The Nucleon Masses}
\label{sec:masses}

\subsection{The Infinite-Volume Limit}

\noindent For purposes of setting notation, we will begin by reviewing the
derivation of leading terms in the chiral expansion of the nucleon mass.
The relevant leading Baryon mass operators in two-flavor $\chi$PT are
\begin{eqnarray}
{\cal L} & = & 
i\overline{N} v\cdot {\cal D} N
\ +\ 2\alpha_M \overline{N}{\cal M}_+ N
\ +\ 2\sigma_M \overline{N} N\ 
{\rm tr}\left[{\cal M}_+\right]
-
i \overline{ T}^\mu v\cdot {\cal D} T_\mu
\ +\ 
\Delta\ \overline{T}^\mu T_\mu
\ \ ,
\label{eq:freeQCD}
\end{eqnarray}
where the chirally-invariant mass operator is 
${\cal M}_+={1\over 2}\left(\xi^\dagger m_q\xi^\dagger + \xi m_q\xi\right)$,
with $m_q={\rm diag}(m_u,m_d)$, and $\xi=\exp{(i\pi_a\tau_a/f)}$
is the usual two-flavor Goldstone matrix.
The relevant leading axial operators are
\begin{eqnarray}
{\cal L} & = & 
2 g_A\  \overline{N} S^\mu  A_\mu N
\ +\ 
g_{\Delta N}\ 
\left[\ 
\overline{T}^{abc,\nu}  A^d_{a,\nu}\,  N_b \, \epsilon_{cd} 
\ +\ {\rm h.c.}
\ \right] \ .
\label{eq:intsQCD}
\end{eqnarray}

\begin{figure}[!ht]
\centerline{{\epsfxsize=4.0in \epsfbox{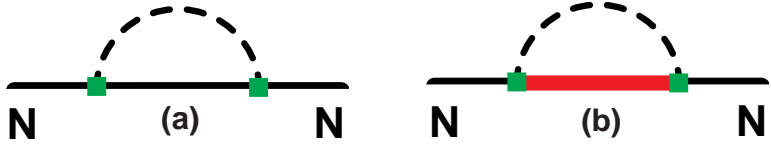}}} 
\vskip 0.15in
\noindent
\caption{\it 
One-loop graphs that give contributions of the form 
$\sim m_q^{3/2}$ to the masses of the proton and neutron.
A solid, thick-solid and dashed line denote a
nucleon, $\Delta$-resonance, and a meson, respectively.
The solid-squares denote an axial coupling.}
\label{fig:masses}
\vskip .2in
\end{figure}

The mass of the $i$-th nucleon has a chiral expansion of the form
\begin{eqnarray}
M_i & = & M_0(\mu)\ -\ M_i^{(1)}(\mu)\ -\ M_i^{(3/2)}(\mu)\ +\ ...
\ \ \ ,
\label{eq:massexp}
\end{eqnarray}
where a term $M_i^{(\alpha)}$ denotes a contribution of order
$m_q^\alpha$, and $i=p,n$. The nucleon mass is dominated by a term in
the $\chi$PT Lagrangian, $M_0$, that is independent of $m_q$.
Here $\Delta$, the $\Delta$-nucleon mass splitting, is assumed to be of the 
same chiral order as the pion mass~\cite{JM,HHK}.
Each of the contributions in eq.~(\ref{eq:massexp}) depends upon the
scale chosen to renormalize the theory.  
While to ${\cal O}\left(m_q\right)$, the objects $M_0$
and $ M_i^{(1)}$ are scale independent, at one-loop level, ${\cal O}(m_q^{3/2})$, they are
scale dependent.  The leading dependence upon $m_q$,
occurring at ${\cal O}\left(m_q\right)$, is due to the operators in
eq.~(\ref{eq:freeQCD}) with coefficients $\alpha_M$
and $\sigma_M$. The leading non-analytic
dependence upon $m_q$ arises from the one-loop diagrams shown in
Fig.~\ref{fig:masses}.

In isospin-symmetric QCD~\footnote{The nucleon masses, including strong isospin breaking, may be obtained by taking the QCD limit
of the partially-quenched expressions given in Ref.~\cite{BSn}.}, with $m_u,m_d\rightarrow\overline{m}$, 
one finds the nucleon mass at one-loop order in the
chiral expansion~\cite{J92},
\begin{eqnarray}
M_N & = & M_0(\mu) - 2 \overline{m}\left( \alpha_M+2\sigma_M\right)(\mu)
- {1\over 8\pi f^2}\left[\ {3\over 2} g_A^2 m_\pi^3
\ +\ {4 g_{\Delta N}^2\over 3\pi} F(m_\pi,\Delta,\mu )\ \right]
\ \ \ ,
\label{eq:massesQCDisolimit}
\end{eqnarray}
where
\begin{eqnarray}
F (m,\Delta,\mu) & = & 
\left(m^2-\Delta^2\right)\left(
\sqrt{\Delta^2-m^2} \log\left({\Delta -\sqrt{\Delta^2-m^2+i\epsilon}\over
\Delta +\sqrt{\Delta^2-m^2+i\epsilon}}\right)
-\Delta \log\left({m^2\over\mu^2}\right)\ \right)
\nonumber\\
& - & {1\over 2}\Delta m^2 \log\left({m^2\over\mu^2}\right)
\ \ \ ,
\label{eq:massfun}
\end{eqnarray}
and $\Delta$ is the $\Delta$-Nucleon mass splitting. Here we have used dimensional regularization (dim reg) with 
$\overline{\rm MS}$ to define the divergent loop integrals and $M_0(\mu)$ and $M^{(1)}(\mu)$.

\subsection{Finite-Size Corrections}

\noindent In the infinite-volume limit, the nucleon mass may be written as
\begin{eqnarray}
M_N & = & M_0({\cal R}) - 2 \overline{m}\left( \alpha_M+2\sigma_M\right)({\cal R})
\ -i\frac{9 g_A^2}{2f^2}\ {\cal I}_{\cal R} (\infty,0) \ -i\frac{4g_{\Delta N}^2}{f^2}\ {\cal I}_{\cal R} (\infty,\Delta) \ ,
\label{eq:massgeneral}
\end{eqnarray}
where 
\begin{eqnarray}
{\cal I}_{\cal R} (\infty,\Delta)\ &=&\  
-{1\over 3}\int_{\cal R} \frac{d^4k}{(2\pi )^4}\ {{{\vec k}^2}\over{({k_0}-\Delta-i\epsilon)(k_0^2-{\vec k}^2-m_\pi^2+i\epsilon)}}\ .
\label{eq:Idefined}
\end{eqnarray}
Here ${\cal R}$ denotes a choice of ultraviolet regulator and a 
renormalization scheme~\footnote{In dim reg with $\overline{\rm MS}$,
\begin{eqnarray}
{\cal I}_{\overline{\rm MS}} (\infty,\Delta)\ 
& =&\ -i{{1}\over{24\pi^2}}\ F (m_\pi,\Delta,\mu) \ , 
\nonumber
\label{eq:IdefinedMS}
\end{eqnarray}
where $F (m_\pi,0,\mu)=\pi m_\pi^3$, which recovers the results of eq.~(\ref{eq:massesQCDisolimit}).}.

In a spatial box of size $L$,  ${\cal I}_{\cal R}$ generalizes to
\begin{eqnarray}
{\cal I}_{\cal R}(L,\Delta)
\ & =& \ i{1\over 3}\ \left(\  
{1\over L^3}\ \sum_{\vec k}^{\cal R}\ \int \frac{dk_4}{(2\pi )}\ {{{\vec k}^2}\over{({ik_4}-\Delta)(k_4^2+{\vec k}^2+m_\pi^2)}} \right) \ ,
\end{eqnarray}
where we have rotated the integral to Euclidean space and accounted for the quantization of the momentum levels due to the periodic
boundary conditions. Feynman parameterizing and explicitly evaluating the $k_4$ integration lead to
\begin{eqnarray}
{\cal I}_{\cal R}(L,\Delta)
\ & =& -i{1\over 6} \int_0^\infty d\lambda\ 
\left(\ {1\over L^3}\ \sum_{\vec k}^{\cal R}\ {{{\vec k}^2 }\over{[{\vec k}^2+\beta_\Delta^2]^{3/2}}}\ \right)\ ,
\end{eqnarray}
where $\beta_\Delta^2\equiv\lambda^2 + 2\lambda \Delta + m_\pi^2$.
We can now write the finite-size corrections to ${\cal I}_{\cal R}$ as
\begin{eqnarray}
\delta_L{\cal I}(\Delta) \ &\equiv& \ {\cal I}_{\cal R}(L,\Delta)\ -\  {\cal I}_{\cal R}(\infty,\Delta) \nonumber \\
&=&   -{i\over 6} \int_0^\infty d\lambda \left[  
\delta_L \left({{1}\over{[{\vec k}^2+\beta_\Delta^2]^{1/2}}}  \right)
 - \beta_\Delta^2  \delta_L \left( {{1}\over{[{\vec k}^2+\beta_\Delta^2]^{3/2}}}  \right) \right] \ ,
\label{eq:deltaIsubLgeneral}
\end{eqnarray}
where $\delta_L \left(f(|{\vec k}|)\right)$ is defined in eq.~(\ref{eq:app1}) of the appendix.
Notice that $\delta_L{\cal I}(\Delta)$ is a purely infrared quantity and, as such, is independent of ${\cal R}$\footnote{This
implies that finite-volume effects should be independent of the lattice spacing $a$, which appears implicitly as the ultraviolet cutoff
$\pi/a$ in all sums and integrals.}.
Using eqs.~(\ref{eq:massgeneral}) and (\ref{eq:deltaIsubLgeneral}), the finite-size corrections to the nucleon mass may then be expressed as
\begin{eqnarray}
\delta_L M_N\ \equiv \ 
M_N(L)\ -\ M_N(\infty)  \ =\ -i\ \frac{9 g_A^2}{2f^2}\ \delta_L{\cal I}(0) \ -\ i\  \frac{4g_{\Delta N}^2}{f^2}\ \delta_L{\cal I}(\Delta ) \ .
\label{eq:exactmassformula}
\end{eqnarray}
Using the ``master'' formula, eq.~(\ref{eq:masterformula}), derived in the appendix, we find
\begin{eqnarray}
\delta_L{\cal I}(\Delta )\ &=&\  
-{i\over 12\pi^2}\ {\cal K}(\Delta) \ ,
\label{eq:delILfinalform}
\end{eqnarray}
where
\begin{eqnarray}
{\cal K}(\Delta) &\equiv&\  
\int_0^\infty d\lambda\ \beta_\Delta\ \sum_{{\vec n}\neq0}\ \left[ \ 
(L\,|{\vec n}|)^{-1} K_1(L\,\beta_\Delta\,|{\vec n}|)\ -\ \beta_\Delta\  K_0(L\,\beta_\Delta\,|{\vec n}|)) \ \right] \ .
\label{eq:defined} 
\end{eqnarray}
\begin{figure}[ht]
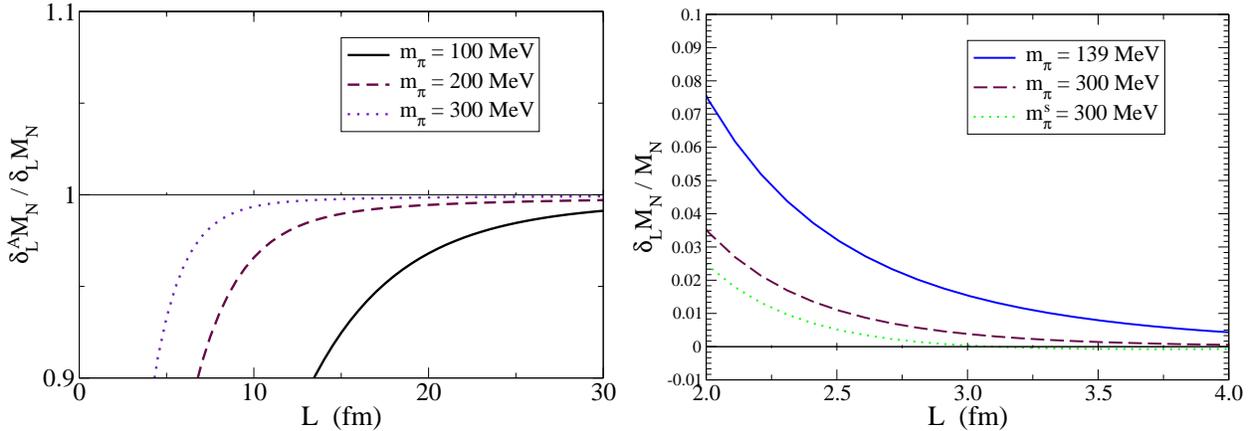

\vskip 0.4in
\centerline{{\epsfxsize=3.2in \epsfbox{Asymovertrue3.eps}}\hskip0.2cm{\epsfxsize=3.2in \epsfbox{MNvsLlinear4.eps}}} 
\caption{\it Left panel:
The ratio of the asymptotic formula, eq.~(\ref{eq:lowenergytheoremasym}), over the exact formula, eq.~(\ref{eq:exactmassformula2compact}),
as a function of $L$ for various values of $m_\pi$.
The solid, dashed and dotted lines correspond to
$m_\pi=100~{\rm MeV}$, $200~{\rm MeV}$ and $300~{\rm MeV}$,
respectively. Right panel: The ratio of the nucleon mass size dependence to the (infinite volume) nucleon mass vs. $L$.
The solid and dashed lines correspond to {\rm QCD} with $m_\pi=139~{\rm MeV}$ and $300~{\rm MeV}$, respectively.
The dotted line corresponds to {\rm PQQCD} with $m_\pi=139~{\rm MeV}$ and $m_\pi^s=300~{\rm MeV}$.}
\label{fig:asymratio}
\end{figure}
Here $K_n(z)$ is a modified Bessel function of the second kind. With $\Delta=0$ the integral over $\lambda$ can be carried
out explicitly (see Appendix) and one has
\begin{eqnarray}
{\cal K}(0) \ &=&\  
-{\pi\over 2}\ m_\pi^2\ 
\sum_{{\vec n}\neq0}\ 
(L\,|{\vec n}|)^{-1} \exp(-L\,|{\vec n}|\, m_\pi)\ .
\label{eq:delzeronew} 
\end{eqnarray}
Notice that this function contains no power-law corrections.
Finally, we have
\begin{eqnarray}
\delta_L M_N\ = \ -\frac{3g_A^2}{8\pi^2f^2}\ {\cal K}(0) \ &-&\ 
\frac{g_{\Delta N}^2}{3\pi^2f^2}\ {\cal K}(\Delta) \ .
\label{eq:exactmassformula2compact}
\end{eqnarray}
This is the exact formula for the finite-size corrections to the nucleon mass at leading order in baryon $\chi$PT.
In Fig.~\ref{fig:asymratio} (right panel), the ratio 
of the nucleon mass size dependence to the (infinite volume) nucleon mass has been plotted against the box size $L$
for various pion masses. The solid and dashed lines correspond to the {\rm QCD} formula of eq.~(\ref{eq:exactmassformula2compact}) 
with $m_\pi=139~{\rm MeV}$ and $300~{\rm MeV}$, respectively. The dotted line corresponds to {\rm PQQCD} in the isospin 
limit taken from eq.~(\ref{eq:NucleonmassPQisolimit}) with $m_\pi=139~{\rm MeV}$ and $m_\pi^s=300~{\rm MeV}$.
We use the parameter set: $f=132~{\rm MeV}$, $g_A=1.26$ and $g_{\Delta N}=1.4$.

\subsection{The Asymptotic Limit}

\noindent Using eqs.~(\ref{eq:delzeronew}) and (\ref{eq:kdelasym}), in the large-$L$ expansion 
one has
\begin{eqnarray}
\delta^A_L M_N\ = \ \left(\ \frac{9 g_A^2 {m_\pi^2}}{8\pi f^2}\ +\ 
\frac{4g_{\Delta N}^2 m_\pi^{5/2}}{(2\pi)^{3/2}f^2}\ {1\over{\Delta\; L^{1/2}}}\ \right) \ {1\over{L}} \exp{(-m_\pi L)}  \ ,
\label{eq:lowenergytheoremasym}
\end{eqnarray}
where $\delta_L M_N-\delta^A_L M_N={\cal O}(\exp{(-m_\pi L)}/L^{5/2})$. 
In the $M_N\rightarrow\infty$ limit, the leading term in the large-$L$ expansion is in agreement
with Ref.~\cite{Khan:2003cu} and in disagreement with Ref.~\cite{Luscher}~\footnote{For a detailed discussion of this
disagreement, see Ref.~\cite{Khan:2003cu}.}.
Fig.~\ref{fig:asymratio} (left panel) plots the ratio $\delta^A_L M_N/\delta_L M_N$ as a function of $L$ for various pion masses. 
Clearly the utility
of eq.~(\ref{eq:lowenergytheoremasym}) is purely aesthetic; even with heavy
pions, the asymptotic formula, eq.~(\ref{eq:lowenergytheoremasym}), is not accurate for $L< 10~{\rm fm}$.
This points to the importance of exponential corrections; for $m_\pi L \gsim 1$  convergence of the momentum sums requires
keeping terms with $|{\vec n}|>1$, {\it i.e.} one must include corrections of ${\cal O}(\exp{(-|{\vec n}|m_\pi L)})$.

\section{The Nucleon Magnetic Moments}
\label{sec:magmoms}

\subsection{The Infinite-Volume Limit}

\noindent With the finite-size corrections for the masses, it is straightforward to get the magnetic moments.
The leading operators contributing to the nucleon magnetic moments are
\begin{eqnarray}
{\cal L} & = & 
{e\over 4 M_N} F_{\mu\nu}\ 
\left(\ 
\mu_0\ \overline{N}\sigma^{\mu\nu} N
\ +\ 
\mu_1\ \overline{N}\sigma^{\mu\nu} 
\tau^3_{\xi+}\ N
\ \right)
\ \ \ ,
\label{eq:magQCDiso}
\end{eqnarray}
where $F_{\mu\nu}$ is the electromagnetic field-strength tensor, 
$M_N$ is the physical value of the nucleon mass,
$\mu_0$ is the isoscalar nucleon magnetic moment,
$\mu_1$ is the isovector nucleon magnetic moment and
$\tau^a_{\xi+}  =  {1\over 2} \left(\ \xi^\dagger\tau^a \xi + \xi\tau^a \xi^\dagger \right)$.

In isospin-symmetric QCD one finds the nucleon magnetic moment matrix at one-loop order in the
\begin{figure}[!ht]
\centerline{{\epsfxsize=4.0in \epsfbox{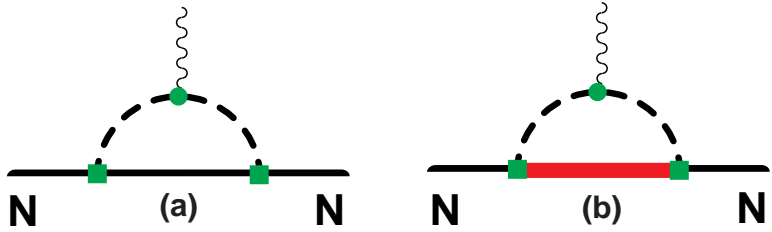}}} 
\vskip 0.15in
\noindent
\caption{\it 
One-loop graphs that contribute to the
proton and neutron magnetic moments.
A solid, thick-solid and dashed line denote a
nucleon, $\Delta$-resonance, and a meson, respectively.
The solid-squares denote an axial coupling and the solid-circles denote
a leading-order electromagnetic interaction.}
\label{fig:magmoms}
\vskip .2in
\end{figure}
chiral expansion~\cite{CP74,JLMS92,MS97}
\begin{eqnarray}
{\hat\mu} & = & \mu_0+\mu_1\ {\hat\tau}_3
- {M_N\over 4\pi f^2}\left[\ g_A^2 \ m_{\pi^+} 
+ {2\over 9}\ g_{\Delta N}^2\  {\cal F}_{\pi^+}\ \right]\ {\hat\tau}_3 \ .
\label{eq:magmomsQCD}
\end{eqnarray}
The scale dependence is left implicit.
The proton and the neutron magnetic moments are the diagonal elements of
${\hat\mu}$. The first term within the brackets is from Fig.~\ref{fig:masses}(a)
while the second term is from Fig.~\ref{fig:masses}(b).
The function $\cf_i={\cal F}(m_i,\Delta,\mu)$ is
\begin{eqnarray}
\pi {\cal F}(m,\Delta,\mu)
& = & \sqrt{\Delta^2-m^2}\log\left({\Delta-\sqrt{\Delta^2-m^2+i\epsilon}
\over \Delta+\sqrt{\Delta^2-m^2+i\epsilon}}\right)
\ -\ \Delta\log\left({m^2\over\mu^2}\right)
\ \ \ .
\label{eq:magfun}
\end{eqnarray}
Here again we have used dim reg with 
$\overline{\rm MS}$. In the limit $\Delta\rightarrow 0$, ${\cal F}(m,0,\mu)=m$.

\subsection{Finite-Size Corrections}

\begin{figure}[ht]
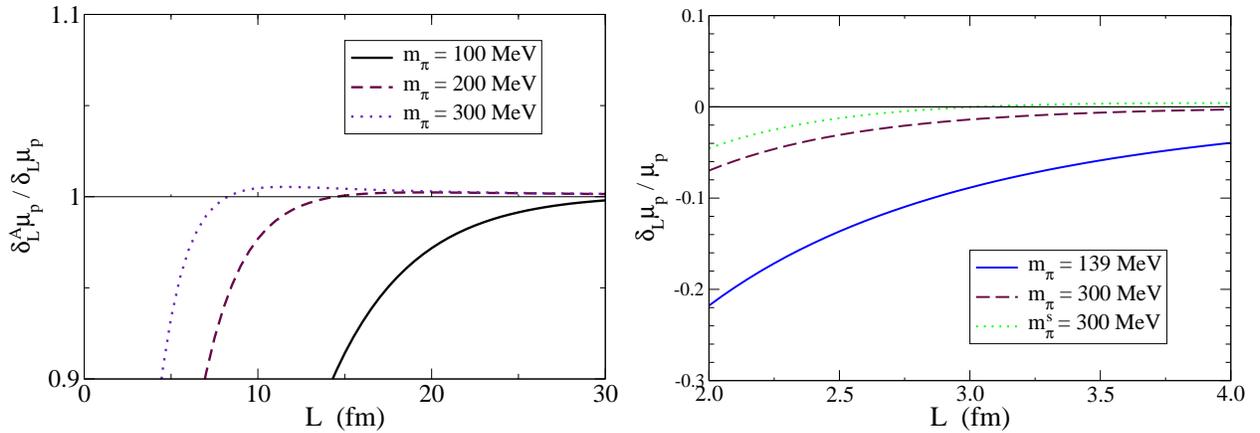

\vskip 0.4in
\centerline{{\epsfxsize=3.2in \epsfbox{MMAsymovertrue6.eps}}\hskip0.2cm{\epsfxsize=3.2in \epsfbox{MMvsLnew7.eps}}} 
\caption{\it Left panel:
The ratio of the asymptotic formula, eq.~(\ref{eq:magmomsfinitesizewithdeltaasymptotic}), over the exact formula, eq.~(\ref{eq:magmomsfinitesizecompact}),
as a function of $L$ for various values of $m_\pi$.
The solid, dashed and dotted lines correspond to
$m_\pi=100~{\rm MeV}$, $200~{\rm MeV}$ and $300~{\rm MeV}$,
respectively. Right panel: The ratio of the nucleon magnetic moment size dependence to the (infinite volume) nucleon nucleon
magnetic moment vs. $L$.
The solid and dashed lines correspond to {\rm QCD} with $m_\pi=139~{\rm MeV}$ and $300~{\rm MeV}$, respectively.
The dotted line corresponds to {\rm PQQCD} with $m_\pi=139~{\rm MeV}$ and $m_\pi^s=300~{\rm MeV}$.}
\label{fig:asymratio2}
\end{figure}

\noindent In the infinite-volume limit, the nucleon magnetic moments may be written as
\begin{eqnarray}
{\hat\mu} & = & \mu_0+\mu_1\ {\hat\tau}_3
- {{4 i M_N}\over f^2}\left[\ g_A^2 \ {\cal J}_{\cal R}(\infty,0)
+ {2\over 9}\ g_{\Delta N}^2\  {\cal J}_{\cal R}(\infty,\Delta)\ \right]\ {\hat\tau}_3 \ ,
\label{eq:magmomsQCDwithJs}
\end{eqnarray}
where
\begin{eqnarray}
{\cal J}_{\cal R}(\infty,\Delta)\ =\ \frac{\partial}{\partial m_\pi^2}\ {\cal I}_{\cal R}(\infty,\Delta)\ .
\label{eq:Jdefined}
\end{eqnarray}
Therefore, the finite-size corrections to ${\hat\mu}$ are
\begin{eqnarray}
\delta_L {\hat\mu}\ \equiv\ {\hat\mu}(L) \ -\ {\hat\mu}(\infty) 
 & = & - {{4 i M_N}\over f^2}\left[\ g_A^2 \ \delta_L{\cal J}(0)
+ {2\over 9}\ g_{\Delta N}^2\  \delta_L{\cal J}(\Delta)\ \right]\ {\hat\tau}_3 \ .
\label{eq:magmomsfinitesize}
\end{eqnarray}
Using eqs.~(\ref{eq:delILfinalform}), (\ref{eq:defined}) and (\ref{eq:Jdefined}), and the properties
of modified Bessel functions, it is straightforward to find
\begin{eqnarray}
\delta_{L}{\cal J}(\Delta )\ &=&\ {i\over 24\pi^2}\ {\cal Y}(\Delta) \ ,
\label{eq:delJLfinalform}
\end{eqnarray}
where
\begin{eqnarray}
{\cal Y}(\Delta) \ &\equiv&\  
\int_0^\infty d\lambda\ \sum_{{\vec n}\neq0}\ \left[ \ 
3 K_0(L\,\beta_\Delta\,|{\vec n}|)\ -\ (L\,\beta_\Delta\,|{\vec n}|)\ K_1(L\,\beta_\Delta\,|{\vec n}|)) \ \right] \ .
\label{eq:Ydef}
\end{eqnarray}
With $\Delta=0$ one has
\begin{eqnarray}
{\cal Y}(0) \ &=&\  
-{\pi\over 2}\ m_\pi\ 
\sum_{{\vec n}\neq0}\ \left(\ 1\ -\ 2(L\,|{\vec n}|\, m_\pi)^{-1}\ \right)\ \exp(-L\,|{\vec n}|\, m_\pi)\ .
\label{eq:delzeronewJ} 
\end{eqnarray}
Finally one arrives at
\begin{eqnarray}
\delta_L {\hat\mu}\ 
 & = & {{M_N}\over 6\pi^2 f^2}\left[\ g_A^2 \ {\cal Y}(0)
+ {2\over 9}\ g_{\Delta N}^2\  {\cal Y}(\Delta)\ \right]\ {\hat\tau}_3 \ .
\label{eq:magmomsfinitesizecompact}
\end{eqnarray}
This is the exact formula for the finite-size corrections to the nucleon magnetic moments at leading order in baryon $\chi$PT.
In Fig.~\ref{fig:asymratio2} (right panel), the ratio 
of the proton magnetic moment size dependence to the (infinite volume) magnetic moment has been plotted against the box size $L$.
The solid and dashed lines correspond to the {\rm QCD} formula of eq.~(\ref{eq:magmomsfinitesizecompact}) 
with $m_\pi=139~{\rm MeV}$ and $300~{\rm MeV}$, respectively. The dotted line corresponds to {\rm PQQCD} in the isospin 
limit taken from eq.~(\ref{eq:NucleonmagPQisolimit}) with $m_\pi=139~{\rm MeV}$ and $m_\pi^s=300~{\rm MeV}$.
\subsection{The Asymptotic Limit}

\noindent Again using eqs.~(\ref{eq:delILfinalform}), (\ref{eq:defined}) and (\ref{eq:Jdefined}), together with
eqs.~(\ref{eq:delzeronewJ}) and (\ref{eq:kdelasym}), in the large-$L$ expansion one 
has
\begin{eqnarray}
\delta_L {\hat\mu} 
 & = & - \left(\ {{M_N g_A^2 m_\pi}\over{2\pi f^2}} \left(\ 1 - {2\over{m_\pi L}} \right) + 
\frac{4 M_N g_{\Delta N}^2 m_\pi^{3/2}}{9(2\pi)^{3/2}f^2}\ {1\over{\Delta\; L^{1/2}}} \right) \exp{(-m_\pi L)}  {\hat\tau}_3 + \ldots
\label{eq:magmomsfinitesizewithdeltaasymptotic}
\end{eqnarray}
where the dots denote contributions of ${\cal O}(\exp{(-m_\pi L)}/L^{3/2})$.
Fig.~\ref{fig:asymratio2} (left panel) plots the ratio $\delta^A_L \mu/\delta_L \mu$ as a function of $L$ for various pion masses. 
The curves are similar to those of the nucleon mass in Fig.~\ref{fig:asymratio}, as is the conclusion about the
practical utility of eq.~(\ref{eq:magmomsfinitesizewithdeltaasymptotic}).

\section{Discussion and Conclusion}
\label{sec:conc}

\noindent It is hoped that in the near future lattice (PQ)QCD simulations of
baryon properties will encounter the chiral regime, where the quark masses are
sufficiently small to allow a meaningful chiral expansion in quark masses, box size
and lattice spacing. It is likely that this regime has been encountered in recent 
work on heavy-meson systems~\cite{Lepage}. 

The results of this paper, together with the results of Refs.~\cite{BSn}
and \cite{Beane:2003xv}, give the dependence of the nucleon masses and magnetic moments on the
sea and valence quark masses {\it and} on the lattice spacing, $a$,
and size, $L$, to leading order in the chiral expansion.  We
eagerly await lattice (PQ)QCD simulations within the chiral regime
where this parameter space may be fruitfully explored.

\acknowledgments

\noindent 
I would like to thank David Lin and Martin Savage for helpful discussions.
This work was partly supported by DOE contract DE-AC05-84ER40150, under which the
Southeastern Universities Research Association (SURA) operates the
Thomas Jefferson National Accelerator Facility.

\appendix{}
\section{Finite Size Corrections}
\label{sec:app1}

\subsection{The Master Formula}

\noindent We wish to evaluate
\begin{eqnarray}
\delta_L \left(\ {{1}\over{[{\vec l}^2+{\cal M}^2]^{m}}}\ \right) 
\ & \equiv &\ {1\over L^3} \sum_{\vec l}\ {{1}\over{({\vec l}^2+{\cal M}^2)^m}}- 
\int \frac{d^3l}{(2\pi )^3}\ {{1}\over{({\vec l}^2+{\cal M}^2)^m}}\ \ .
\label{eq:app1}
\end{eqnarray}
As this difference is ultraviolet finite, we omit the label denoting the scheme dependence
of the individual sum and integral. This expression has been evaluated in many 
places~\footnote{References that have been of use to the author include 
Refs.~\cite{Elizalde:1997jv,Becirevic:2003wk,savagenotes}.}.
Using the identity
\begin{eqnarray}
D^{-m}\ =\ {1\over{\Gamma(m)}}\int_0^\infty  d\eta \eta^{m-1} e^{-\eta D} 
\label{eq:app2}
\end{eqnarray}
one finds
\begin{eqnarray}
\hspace{-.2cm}
\delta_L \left(\ {{1}\over{[{\vec l}^2+{\cal M}^2]^{m}}}\ \right) = 
\frac{1}{(4\pi)^{3/2}\Gamma(m)} \int_0^\infty  d\eta \eta^{m-5/2} e^{-\eta {{\cal M}^2}} \left[
\frac{(4\pi\eta)^{3/2}}{L^3} \sum_{\vec l} e^{-\eta{\vec l}^{\;2}}- 1  \right] .
\label{eq:app3}
\end{eqnarray}
Expressing the momentum as ${\vec l}=2\pi{\vec n}/L$ and using the Jacobi identity~\cite{Elizalde:1997jv},
\begin{eqnarray}
{\cal S}(z)\ \equiv\ \sum_{n=-\infty}^{\infty} e^{-z n^2} \ , \qquad {\cal S}(z)\ =\ \sqrt{{{\pi}\over z}}\  {\cal S}({{\pi^2}\over z}) \ ,
\label{eq:app5}
\end{eqnarray}
leads to
\begin{eqnarray}
\delta_L \left(\ {{1}\over{[{\vec l}^2+{\cal M}^2]^{m}}}\ \right) 
\ & =&  \frac{1}{(4\pi)^{3/2}\Gamma(m)} \sum_{{\vec n}\neq 0}\int_0^\infty  d\eta \eta^{m-5/2} e^{-\eta {{\cal M}^2}} 
e^{-{L^2{\vec n}^2}/{4\eta}} \ .
\label{eq:app6}
\end{eqnarray}
Performing the integral over $\eta$ then gives the ``master'' formula
\begin{eqnarray}
\delta_L \left(\ {{1}\over{[{\vec l}^2+{\cal M}^2]^{m}}}\ \right) 
\ & =& \
\frac{2^{-\frac{1}{2}-m}\,{\cal M}^{3 - 2\,m}}{{\pi }^{3/2}\, \Gamma(m)}
\sum_{{\vec n}\neq 0}\ 
{\left( L\,{\cal M}\,|{\vec n}| \right) }^{-\frac{3}{2}+m}\,  K_{\frac{3}{2} - m}(L\,{\cal M}\,|{\vec n}|) \ ,
\label{eq:masterformula}
\end{eqnarray}
where $K_n(z)$ is a modified Bessel function of the second kind. 

A well-chosen change of integration variable and 
the properties of modified Bessel functions allow one to write eq.~(\ref{eq:defined}) as
\begin{eqnarray}
{\cal K}(\Delta)  &=&
\sum_{{\vec n}\neq0}\ (L\,|{\vec n}|)^{-1}\ {1\over L}\ {d\over dL}\ \left( L^2 
\int_{m_\pi}^\infty d\xi\ \xi^{2}\ (\xi^2-m_\pi^2+\Delta^2)^{-1/2}\ K_1(L\,\xi\,|{\vec n}|)\ \right).
\label{eq:usefulformula}
\end{eqnarray}
We find no useful simplification of this formula in the general case. With $\Delta=0$ one directly finds
\begin{eqnarray}
{\cal K}(0) \ &=&\  
-\sqrt{\pi\over 2}\ m_\pi^3\ \sum_{{\vec n}\neq0}\ 
(L\,|{\vec n}|\,m_\pi)^{-{1\over 2}} K_{1\over 2}(L\,|{\vec n}|\, m_\pi)\ ,
\end{eqnarray}
which gives eq.~(\ref{eq:delzeronew}).

\subsection{The Asymptotic Limit}

\noindent In the large-$L$ limit, using the expansion of the modified Bessel function for
large argument, one finds from eq.~(\ref{eq:usefulformula}),
\begin{eqnarray}
{\cal K}(\Delta)  &=& 3\sqrt{2\pi}\ {1\over L^2}\ {d\over dL}\ \left( L^{3/2} 
\int_{m_\pi}^\infty d\xi\ \xi^{3/2}\ (\xi^2-m_\pi^2+\Delta^2)^{-1/2}\ \exp(-L\,\xi)\ \right)+\ldots 
\label{eq:usefulformula2}
\end{eqnarray}
where the dots denote contributions of ${\cal O}(\exp{(-m_\pi L)}/L^{5/2})$.
Observe that one can expand the integrand in powers of $\alpha^2\equiv \Delta^2-m_\pi^2$,
\begin{eqnarray}
\int_{m_\pi}^\infty d\xi\ \xi^{\ell/2}\ (\xi^2-m_\pi^2+\Delta^2)^{-1/2}\ e^{-L\ \xi} &=&
\sum_{n=0}^\infty\  \left(\matrix{-\frac{1}{2}\cr n}\right)\ \alpha^{2 n}\ \int_{m_\pi}^\infty d\xi\ \xi^{\ell/2-1-2n}\ e^{-L\ \xi} 
\nonumber \\
&=& m_\pi^{\ell/2-1}\ {1\over{L}}\ e^{-m_\pi L}\ \sum_{n=0}^\infty\  \left(\matrix{-\frac{1}{2}\cr n}\right)\ \frac{\alpha^{2 n}}{m_\pi^{2 n}}+\ldots
\nonumber \\
&=& m_\pi^{\ell/2}\ {1\over{{\Delta}\;L}}\ \exp{(-L\ m_\pi)}\ +\ \ldots
\end{eqnarray}
where the dots denote contributions of ${\cal O}(\exp{(-m_\pi L)}/L^{2})$.
(Similar technology has been developed in Ref.~\cite{ArLi} in the context of the heavy-meson systems.)
Plugging this back into eq.~(\ref{eq:usefulformula2}) one finds, in the asymptotic limit,
\begin{eqnarray}
{\cal K}(\Delta) \ &=&\  -3\sqrt{2\pi}\ {m_\pi^{5/2}}\ {1\over{ L^{3/2}\Delta}} \exp{(-L\ m_\pi)} \ +\ \ldots\ .
\label{eq:kdelasym}
\end{eqnarray}
where the dots denote contributions of ${\cal O}(\exp{(-m_\pi L)}/L^{5/2})$.

\section{Partially-Quenched QCD}
\label{sec:app2}

\subsection{Nucleon Masses}

\noindent We work in PQQCD including isospin breaking, but with electromagnetism turned 
off~\footnote{For details we refer the reader to Ref.~\cite{BSn}.}.
The Lagrangian describing the interactions of 
${\cal B}_{ijk}$ (containing the nucleon) and ${\cal T}_{ijk}$ (containing $\Delta$), which transform in the 
${\bf 70}$ and ${\bf 44}$ of $SU(4|2)_V$, respectively,
with the pseudo-Goldstone bosons at leading-order in the chiral expansion
is~\cite{LS96}
\begin{eqnarray}
{\cal L} & = & 
2\alpha\ \left(\overline{\cal B} S^\mu {\cal B} A_\mu\right)
\ +\ 
2\beta\ \left(\overline{\cal B} S^\mu A_\mu {\cal B} \right)
\ +\  
\sqrt{3\over 2}{\cal C} 
\left[\ 
\left( \overline{\cal T}^\nu A_\nu {\cal B}\right)\ +\ 
\left(\overline{\cal B} A_\nu {\cal T}^\nu\right)\ \right]
\  .
\label{eq:PQints}
\end{eqnarray}
Here the axial-vector field $A_\mu$ is a six-by-six matrix.
Matching to the QCD effective Lagrangian of eq.~(\ref{eq:intsQCD}) and to the
additional operator
\begin{eqnarray}
{\cal L} & = & 
g_1\overline{N} S^\mu N\ {\rm tr}
\left[\ A_\mu\ \right]
\ ,
\label{eq:intsQCDextra}
\end{eqnarray}
one finds that at tree-level,
\begin{eqnarray}
\alpha & = & {4\over 3} g_A\ +\ {1\over 3} g_1
\ \ \ ,\ \ \ 
\beta \ =\ {2\over 3} g_1 - {1\over 3} g_A
\ \ \ ,\ \ \ 
{\cal C} \ =\ -g_{\Delta N}\ .
\label{eq:axrels}
\end{eqnarray}

The finite-size corrections to the proton mass are given by
\begin{eqnarray}
& & \delta_L M_p \ =\   -{1\over 8\pi f^2}\left(\ 
{2g_A^2\over{3\pi}}\left[\ 
{\cal K}(m_{uu},0)+{\cal K}(m_{ud},0)+2{\cal K}(m_{ju},0)+ 2{\cal K}(m_{lu},0) + 3 {\cal G}_{\eta_u , \eta_u}(0)
\ \right]
\right.\nonumber\\
& & \left. \quad
\ +\ {g_1^2\over {6\pi}}\left[\ 
{\cal K}(m_{uu},0) - 5{\cal K}(m_{ud},0) + 
3{\cal K}(m_{jd},0) + 2{\cal K}(m_{ju},0) + 3{\cal K}(m_{ld},0)
+ 2{\cal K}(m_{lu},0) 
\right.\right. \nonumber\\
& & \left.\left. \quad 
 + 3 {\cal G}_{\eta_u , \eta_u}(0)
+ 6 {\cal G}_{\eta_u , \eta_d}(0) + 3 {\cal G}_{\eta_d , \eta_d}(0)
\  \right]
\right.\nonumber\\
& & \left. \quad
\ +\ {2g_A g_1\over{3\pi}}\left[\ 
{\cal K}(m_{ju},0) + {\cal K}(m_{lu},0) - {\cal K}(m_{ud},0) + 2{\cal K}(m_{uu},0) + 3 {\cal G}_{\eta_u , \eta_d}(0)
+ 3 {\cal G}_{\eta_u , \eta_u}(0)
\  \right]
\right.\nonumber\\
& & \left. 
+ {2g_{\Delta N}^2\over 9\pi}\left[\ 
5{\cal K}(m_{ud},\Delta)  + {\cal K}(m_{uu},\Delta) + {\cal K}(m_{ju},\Delta) 
+ {\cal K}(m_{lu},\Delta) + 2{\cal K}(m_{jd},\Delta) + 2{\cal K}(m_{ld},\Delta)
\right.\right. \nonumber\\
& & \left.\left. \
+ 2 {\cal G}_{\eta_d , \eta_d}(\Delta) +2 {\cal G}_{\eta_u , \eta_u}(\Delta)
- 4 {\cal G}_{\eta_u , \eta_d}(\Delta)
\ \right]\ \right)
\ \ \ ,
\label{eq:pmass}
\end{eqnarray}
where ${\cal G}_{\eta_a ,\eta_b}(\Delta)\equiv
{\cal H}_{\eta_a\eta_b}({\cal K}(m_{\eta_a},\Delta),{\cal K}(m_{\eta_b},\Delta),{\cal K}(m_X,\Delta))$,
${\cal H}_{\eta_a\eta_b}$ is given by
\begin{eqnarray}
& & {\cal H}_{\eta_a\eta_b}( A, B, C) \ =\  
-{1\over 2}\left[\ 
{(m_{jj}^2-m_{\eta_a}^2)(m_{ll}^2-m_{\eta_a}^2)\over 
(m_{\eta_a}^2-m_{\eta_b}^2)(m_{\eta_a}^2-m_X^2)}\  A
-
{(m_{jj}^2-m_{\eta_b}^2)(m_{ll}^2-m_{\eta_b}^2)\over 
(m_{\eta_a}^2-m_{\eta_b}^2)(m_{\eta_b}^2-m_X^2)}\  B
\right.\nonumber\\ & & \left.\qquad
\ +\ 
{(m_X^2-m_{jj}^2)(m_X^2-m_{ll}^2)\over 
(m_X^2-m_{\eta_a}^2)(m_X^2-m_{\eta_b}^2)}\  C
\ \right]
\ \ \ ,
\label{eq:HPsdef}
\end{eqnarray}
the mass, $m_X$, is given by 
$m_X^2 = {1\over 2}\left( m_{jj}^2+ m_{ll}^2 \right)$, and ${\cal K}(m_\pi,\Delta)$ is defined
in eq.~(\ref{eq:defined}) (where now the $m_\pi$ dependence is made explicit). Note that
$m_{ab}$ refers to the Goldstone-boson mass with quark content $a$ and $b$ (hence $m_{\pi^\pm}=m_{ud}$, {\it etc.}); 
$j$ and $l$ label the sea quark masses.

The finite-size corrections to the neutron mass are given by
\begin{eqnarray}
& & \delta_L M_n \ =\   -{1\over 8\pi f^2}\left(\ 
{2g_A^2\over{3\pi}}\left[\ 
{\cal K}(m_{dd},0)+{\cal K}(m_{ud},0)+2{\cal K}(m_{jd},0)+ 2{\cal K}(m_{ld},0) + 3 {\cal G}_{\eta_d , \eta_d}(0)
\ \right]
\right.\nonumber\\
& & \left. \quad
\ +\ {g_1^2\over {6\pi}}\left[\ 
{\cal K}(m_{dd},0) - 5{\cal K}(m_{ud},0) + 2{\cal K}(m_{jd},0) +
3{\cal K}(m_{ju},0) + 2{\cal K}(m_{ld},0) + 3{\cal K}(m_{lu},0)
\right.\right. \nonumber\\
& & \left.\left. \quad 
 + 3 {\cal G}_{\eta_u , \eta_u}(0)
+ 6 {\cal G}_{\eta_u , \eta_d}(0) + 3 {\cal G}_{\eta_d , \eta_d}(0)
\  \right]
\right.\nonumber\\
& & \left. \quad
\ +\ {2g_A g_1\over{3\pi}}\left[\ 
2{\cal K}(m_{dd},0) + {\cal K}(m_{jd},0) + {\cal K}(m_{ld},0) - {\cal K}(m_{ud},0) +  3 {\cal G}_{\eta_d , \eta_d}(0)
+ 3 {\cal G}_{\eta_u , \eta_d}(0)
\  \right]
\right.\nonumber\\
& & \left. 
+ {2g_{\Delta N}^2\over 9\pi}\left[\ 
5{\cal K}(m_{ud},\Delta)  + {\cal K}(m_{dd},\Delta) + {\cal K}(m_{jd},\Delta) 
+ {\cal K}(m_{ld},\Delta) + 2{\cal K}(m_{ju},\Delta) + 2{\cal K}(m_{lu},\Delta)
\right.\right. \nonumber\\
& & \left.\left. \quad
+ 2 {\cal G}_{\eta_d , \eta_d}(\Delta) +2 {\cal G}_{\eta_u , \eta_u}(\Delta)
- 4 {\cal G}_{\eta_u , \eta_d}(\Delta)
\ \right]\ \right)
\ \ \ .
\label{eq:nmass}
\end{eqnarray}
In the isospin limit, one has
\begin{eqnarray}
\delta_L M_N \ &=&\  -{g_A^2\over{24\pi^2f^2}}\left[\ {\cal K}(m_\pi,0)+8{\cal K}(m^s_\pi,0) \ \right]\
-\ {g_{\Delta N}^2\over 6\pi^2f^2}\left[\ {\cal K}(m_\pi,\Delta)  + {\cal K}(m^s_\pi,\Delta)\  \right]
\nonumber\\
&& \qquad\qquad +\ 
{g_1\over {24 \pi^2 f^2}}\ \left( 5g_1 + 4g_A \right)\ \left[\ {\cal K}(m_\pi,0)-{\cal K}(m^s_\pi,0)\ \right]
\ \ \ ,
\label{eq:NucleonmassPQisolimit}
\end{eqnarray}
where we have used the fact that ${\cal G}_{\eta_d , \eta_d}(\Delta)\rightarrow -{1\over 2}{\cal K}(m_\pi,\Delta)$
in the isospin limit. Here $m_\pi^s$ denotes the mass of a pion made of one valence quark and one sea quark.
These expressions further collapse down to isospin-symmetric QCD in the limit  $m^s_\pi\rightarrow m_\pi$.

\subsection{Nucleon Magnetic Moments}

\noindent  The most general charge matrix whose matrix elements reduce
to those of QCD (keeping the valence-quark charges fixed) is~\cite{BSn}
\begin{eqnarray}
{\cal Q}^{(PQ)} & = & 
{\rm diag}\left(\ +{2\over 3}\ ,\  -{1\over 3}
\ ,\ q_j\ ,\  q_l\ ,\  q_j\ ,\  q_l\ \right)
\ \ \ .
\label{eq:PQcharge}
\end{eqnarray}

The finite-size corrections to the proton magnetic moment in PQQCD are
\begin{eqnarray}
\delta_L \mu_p & = & -{M_N\over 6\pi^2 f^2}\ \left(\ 
{g_A^2\over 9}\left[\ 4 \;{\cal Y}(m_{uu},0) - 5 \;{\cal Y}(m_{ud},0) - 4 \;{\cal Y}(m_{ju},0) - 4 \;{\cal Y}(m_{lu},0)\ \right]
\right. \nonumber\\ & & \left.
\ +\ 
{2 g_1 g_A\over 9}\left[\ \;{\cal Y}(m_{ud},0) + \;{\cal Y}(m_{uu},0) - \;{\cal Y}(m_{ju},0) - \;{\cal Y}(m_{lu},0)\ \right]
\right. \nonumber\\ & & \left.
\ +\ 
{g_1^2\over 36}\left[\ 
\;{\cal Y}(m_{ud},0) + 4 \;{\cal Y}(m_{uu},0) - 3 \;{\cal Y}(m_{dd},0) + 3 \;{\cal Y}(m_{jd},0) 
\right.\right. \nonumber\\
& & \left.\left. \qquad\qquad
- 4 \;{\cal Y}(m_{ju},0) + 3 \;{\cal Y}(m_{ld},0) 
- 4 \;{\cal Y}(m_{lu},0)\
\right]
\right. \nonumber\\
& & \left.
+
q_j\left({2 g_A^2\over 3}\left[\;{\cal Y}(m_{ju},0)-\;{\cal Y}(m_{uu},0)\right]
+{g_1 g_A\over 3}\left[\;{\cal Y}(m_{ju},0)-\;{\cal Y}(m_{uu},0)\right]
\right.\right. \nonumber\\
& & \left.\left. \qquad\qquad
+{g_1^2\over 6}\left[\;{\cal Y}(m_{ju},0) - \;{\cal Y}(m_{uu},0) + {3\over 2} \;{\cal Y}(m_{jd},0) - 
{3\over 2} \;{\cal Y}(m_{ud},0)\right]
\right)
\right. \nonumber\\
& & \left.
+ 
q_l\left({2 g_A^2\over 3}\left[\;{\cal Y}(m_{lu},0)-\;{\cal Y}(m_{ud},0)\right]
+{g_1 g_A\over 3}\left[\;{\cal Y}(m_{lu},0)-\;{\cal Y}(m_{ud},0)\right]
\right.\right. \nonumber\\
& & \left.\left. \qquad\qquad
+{g_1^2\over 6}\left[\;{\cal Y}(m_{lu},0) - \;{\cal Y}(m_{ud},0) + {3\over 2} \;{\cal Y}(m_{ld},0) - 
{3\over 2} \;{\cal Y}(m_{dd},0)\right]
\right)
\right. \nonumber\\
& & \left.
+\ {g_{\Delta N}^2\over 27}\ \left[\ 
\;{\cal Y}(m_{dd},\Delta) - \;{\cal Y}(m_{uu},\Delta) - 6 \;{\cal Y}(m_{ud},\Delta)
- \;{\cal Y}(m_{jd},\Delta) 
\right.\right. \nonumber\\
& & \left.\left.\qquad\qquad\qquad
+ \;{\cal Y}(m_{ju},\Delta) - \;{\cal Y}(m_{ld},\Delta) + \;{\cal Y}(m_{lu},\Delta)
\right. \right.\nonumber\\
& & \left. \left.\qquad
+ {3\over 2}\  q_j\  
\left( \;{\cal Y}(m_{uu},\Delta) + 2 \;{\cal Y}(m_{ud},\Delta) - \;{\cal Y}(m_{ju},\Delta) - 2 \;{\cal Y}(m_{jd},\Delta)\right)
\right.\right. \nonumber\\
& & \left.\left.\qquad
+ {3\over 2} \ q_l\  
\left( \;{\cal Y}(m_{ud},\Delta) + 2 \;{\cal Y}(m_{dd},\Delta) - \;{\cal Y}(m_{lu},\Delta) - 2 \;{\cal Y}(m_{ld},\Delta)\right)
\ \right] \ \right),
\label{eq:pmag}
\end{eqnarray}
where ${\cal Y}(m_\pi,\Delta)$ is defined
in eq.~(\ref{eq:Ydef}) (where now the $m_\pi$ dependence is made explicit).

The finite-size corrections to the neutron magnetic moment are
\begin{eqnarray}
\delta_L \mu_n & = & -{M_N\over 6\pi^2 f^2}\ \left(\ 
{g_A^2\over 9}\left[\ 7 \;{\cal Y}(m_{ud},0) + 2 \;{\cal Y}(m_{ld},0) + 2 \;{\cal Y}(m_{jd},0) - 2 \;{\cal Y}(m_{dd},0)\ \right]
\right. \nonumber\\ & & \left.
\ +\ 
{g_1 g_A\over 9}\left[\ \;{\cal Y}(m_{jd},0) + \;{\cal Y}(m_{ld},0) - \;{\cal Y}(m_{ud},0) - \;{\cal Y}(m_{dd},0)\ \right]
\right. \nonumber\\ & & \left.
\ +\ 
{g_1^2\over 18}\left[\ 
3\;{\cal Y}(m_{uu},0) + 2 \;{\cal Y}(m_{ud},0) -  \;{\cal Y}(m_{dd},0) +  \;{\cal Y}(m_{jd},0) 
\right.\right. \nonumber\\
& & \left.\left. \qquad\qquad
- 3 \;{\cal Y}(m_{ju},0) +  \;{\cal Y}(m_{ld},0) 
- 3 \;{\cal Y}(m_{lu},0)\
\right]
\right. \nonumber\\
& & \left.
+
q_j\left({2 g_A^2\over 3}\left[\;{\cal Y}(m_{jd},0)-\;{\cal Y}(m_{ud},0)\right]
+{g_1 g_A\over 3}\left[\;{\cal Y}(m_{jd},0)-\;{\cal Y}(m_{ud},0)\right]
\right.\right. \nonumber\\
& & \left.\left. \qquad\qquad
+{g_1^2\over 6}\left[\;{\cal Y}(m_{jd},0) - \;{\cal Y}(m_{ud},0) + {3\over 2} \;{\cal Y}(m_{ju},0) - 
{3\over 2} \;{\cal Y}(m_{uu},0)\right]
\right)
\right. \nonumber\\
& & \left.
+ 
q_l\left({2 g_A^2\over 3}\left[\;{\cal Y}(m_{ld},0)-\;{\cal Y}(m_{dd},0)\right]
+{g_1 g_A\over 3}\left[\;{\cal Y}(m_{ld},0)-\;{\cal Y}(m_{dd},0)\right]
\right.\right. \nonumber\\
& & \left.\left. \qquad\qquad
+{g_1^2\over 6}\left[\;{\cal Y}(m_{ld},0) - \;{\cal Y}(m_{dd},0) + {3\over 2} \;{\cal Y}(m_{lu},0) - 
{3\over 2} \;{\cal Y}(m_{ud},0)\right]
\right)
\right. \nonumber\\
& & \left.
+\ {g_{\Delta N}^2\over 54}\ \left[\ 
\;{\cal Y}(m_{dd},\Delta) - 4\;{\cal Y}(m_{uu},\Delta) +9 \;{\cal Y}(m_{ud},\Delta)
- \;{\cal Y}(m_{jd},\Delta) 
\right.\right. \nonumber\\
& & \left.\left.\qquad\qquad\qquad
+ 4\;{\cal Y}(m_{ju},\Delta) - \;{\cal Y}(m_{ld},\Delta) + 4\;{\cal Y}(m_{lu},\Delta)
\right. \right.\nonumber\\
& & \left. \left.\quad
+ {3}\  q_j\  
\left( \;{\cal Y}(m_{ud},\Delta) + 2 \;{\cal Y}(m_{uu},\Delta) - \;{\cal Y}(m_{jd},\Delta) - 2 \;{\cal Y}(m_{ju},\Delta)\right)
\right.\right. \nonumber\\
& & \left.\left.\quad
+ {3} \ q_l\  
\left( \;{\cal Y}(m_{dd},\Delta) + 2 \;{\cal Y}(m_{ud},\Delta) - \;{\cal Y}(m_{ld},\Delta) - 2 \;{\cal Y}(m_{lu},\Delta)\right)
\ \right] \ \right).
\label{eq:nmag}
\end{eqnarray}
In the isospin limit (with $q_j=q_l=0$), one has
\begin{eqnarray}
\delta_L {\hat\mu} \ &=&\ {M_N\over 6\pi^2 f^2}\ \left(\ 
{g_A^2\over{9}}\left[\ {\cal Y}(m_\pi,0)+8{\cal Y}(m^s_\pi,0) \ \right]\ +\ {2g_{\Delta N}^2\over 9}\ {\cal Y}(m_\pi,\Delta) 
\right.\nonumber\\
&& \left. \qquad\qquad -\ {g_1\over {18}}\ \left( g_1 + 8g_A \right)\ \left[\ {\cal Y}(m_\pi,0)-{\cal Y}(m^s_\pi,0)\ \right]
\right)\  {\hat\tau}_3 \ .
\label{eq:NucleonmagPQisolimit}
\end{eqnarray}
These expressions further collapse down to isospin-symmetric QCD in the limit  $m^s_\pi\rightarrow m_\pi$.

\vfill\eject


\begin{thebibliography}{10}

\bibitem{Lepage}
C.T.H.~Davies {\it et al.}  [HPQCD Collaboration],
{\it Phys. Rev. Lett.} {\bf 92}, 022001 (2004).

\bibitem{Jansen}
For a recent discussion, see K.~Jansen,
{\it Nucl. Phys. Proc. Suppl.} {\bf 129-130}, 3 (2004).

\bibitem{Leutwyler:1987ak}
H.~Leutwyler,
{\it Phys. Lett.} {\bf B189}, 197 (1987).

\bibitem{Leutwyler:1992yt}
H.~Leutwyler and A.~Smilga,
Phys.\ Rev.\ D {\bf 46}, 5607 (1992).

\bibitem{Gasser:1987zq}
J.~Gasser and H.~Leutwyler,
{\it Nucl. Phys.} {\bf B307} 763 (1988).

\bibitem{betterways}
P.~Hasenfratz and H.~Leutwyler,
{\it Nucl. Phys.} {\bf B343} 241 (1990).

\bibitem{Luscher}
M.~L\"uscher,
{\it Lecture given at Cargese Summer Inst., Cargese, France, Sep 1-15, 1983}.

\bibitem{Colangelo:2002hy}
G.~Colangelo, S.~D\"urr and R.~Sommer,
{\it Nucl. Phys. Proc. Suppl.} {\bf 119}, 254 (2003).

\bibitem{Colangelo:2003hf}
G.~Colangelo and S.~D\"urr,
{\it Eur. Phys. J.} {\bf C33}, 543 (2004).

\bibitem{S92}
S.R.~Sharpe, 
{\it Phys. Rev.} {\bf D46}, 3146 (1992). 

\bibitem{golter1}
M.F.L.~Golterman and K.-C.~Leung, 
{\it Phys. Rev.} {\bf D56}, 2950 (1997).

\bibitem{Pqqcd2}
M.F.L.~Golterman and K.-C.~Leung,
{\it Phys. Rev.} {\bf D57}, 5703 (1998).

\bibitem{golter3}
 M.F.L.~Golterman and K.-C.~Leung, 
{\it Phys. Rev.} {\bf D58}, 097503 (1998).

\bibitem{Leinweber:2001ac}
D.B.~Leinweber {\it et al.},
{\it Phys. Rev.} {\bf D64}, 094502 (2001).

\bibitem{davidlin}
 C.-J.D.~Lin {\it et al.},
{\it  Nucl. Phys.} {\bf B650}, 301 (2003).

\bibitem{Becirevic:2003wk}
D.~Becirevic and G.~Villadoro,
{\it Phys. Rev.} {\bf D69}, 054010 (2004).

\bibitem{ArLi}
D.~Arndt and C.-J.D.~Lin, 
{\tt hep-lat/0403012}.

\bibitem{AliKhan:2002hz}
A.~Ali Khan {\it et al.}  [QCDSF Collaboration],
{\it Nucl. Phys. Proc. Suppl.}  {\bf 119}, 419 (2003).

\bibitem{AliKhan:2003kb}
A.~Ali Khan {\it et al.}  [QCDSF Collaboration],
{\tt hep-lat/0309133}.

\bibitem{AliKhan:2003rw}
A.~Ali Khan {\it et al.}  [QCDSF and UKQCD Collaborations],
{\tt hep-lat/0312029}.

\bibitem{Khan:2003cu}
A.~Ali Khan {\it et al.},
{\tt hep-lat/0312030}.

\bibitem{Kronfeld:2002pi}
A.S.~Kronfeld,
{\it At the Frontiers of Particle Physics: Handbook of QCD, Chapter 39, Vol. 4}, edited by M. Shifman.
p. 2411-2477; 
{\tt hep-lat/0205021}.

\bibitem{Young:2002cj}
R.D.~Young {\it et al.},
{\it Phys. Rev.} {\bf D66}, 094507 (2002).

\bibitem{hemmweise}
T.R.~Hemmert and W.~Weise,
{\it Eur. Phys. J.} {\bf A15}, 487 (2002).

\bibitem{hemmver}
V.~Bernard, T.R.~Hemmert and U.-G.~Mei\ss ner,
{\it Nucl. Phys.} {\bf A732}, 149 (2004).

\bibitem{Sharpe90}
S.R.~Sharpe,
{\it Nucl. Phys. Proc. Suppl.}  {\bf 17}, 146 (1990).

\bibitem{BG92}
C.~Bernard and M.F.L.~Golterman,
{\it Phys. Rev.}  {\bf D46}, 853 (1992).

\bibitem{LS96}
J.N.~Labrenz and S.R.~Sharpe,
{\it Phys. Rev.}  {\bf D54}, 4595 (1996).

\bibitem{S01a}
M.J.~Savage,
{\it Nucl. Phys.} {\bf A700}, 359 (2002).

\bibitem{Pqqcd1}
S.R.~Sharpe and N.~Shoresh,
{\it Phys. Rev.} {\bf D62}, 094503 (2000);
{\it Nucl. Phys. Proc. Suppl.} {\bf 83}, 968 (2000).

\bibitem{Pqqcd3}
S.R.~Sharpe,
{\it Phys. Rev.} {\bf D56}, 7052 (1997).

\bibitem{Pqqcd4}
C.W.~Bernard and M.F.L.~Golterman,
{\it Phys. Rev.} {\bf D49}, 486 (1994).

\bibitem{SS01}
S.R.~Sharpe and N.~Shoresh,
{\tt hep-lat/0011089};
{\it Phys. Rev.} {\bf D64}, 114510 (2001).

\bibitem{CSn}
J.-W.~Chen and M.J.~Savage,
{\it Phys. Rev.} {\bf D65}, 094001 (2002);
{\it Phys. Rev.} {\bf D66}, 074509 (2002).

\bibitem{BSn}
S.R.~Beane and M.J.~Savage,
{\it Nucl. Phys.} {\bf A709}, 319 (2002)

\bibitem{BSpv}
S.R.~Beane and M.J.~Savage,
{\it Nucl. Phys.} {\bf B636}, 291 (2002).

\bibitem{Leinweber:2002qb}
D.B.~Leinweber,
{\it Phys. Rev.} {\bf D69}, 014005 (2004).

\bibitem{Arndt:2003ww}
D.~Arndt and B.C.~Tiburzi,
{\it Phys. Rev.} {\bf D68}, 094501 (2003); {\bf D68}, 114503 (2003); {\bf D69}, 014501 (2004).

\bibitem{BSnn}
S.R.~Beane and M.J.~Savage,
{\it Phys. Lett.} {\bf B535}, 177 (2002);
{\it Phys. Rev.} {\bf D67}, 054502 (2003).

\bibitem{ABSl}
D.~Arndt, S.R.~Beane and M.J.~Savage,
{\it Nucl. Phys.} {\bf A726}, 339 (2003).

\bibitem{Beane:2003yx}
S.R.~Beane {\it et al.},
{\tt nucl-th/0311027}.

\bibitem{JM}
E.~Jenkins and A.V.~Manohar,
{\it Phys. Lett.} {\bf B255}, 558 (1991).

\bibitem{HHK}
T.R.~Hemmert, B.R.~Holstein and J.~Kambor,
{\it J. Phys.} {\bf G24}, 1831 (1998).

\bibitem{J92}
E.~Jenkins,
{\it Nucl. Phys.} {\bf B368}, 190 (1992);
V.~Bernard, N.~Kaiser and U.-G.~Mei\ss ner,
{\it Z. Phys.} {\bf C60}, 111 (1993).

\bibitem{CP74}D. G.~Caldi and H.~Pagels,
{\it Phys. Rev.} {\bf D10}, 3739 (1974).

\bibitem{JLMS92}
E.~Jenkins {\it et al.},
{\it Phys. Lett.} {\bf B302}, 482 (1993); {\bf B388}, 866 (1996)(E).

\bibitem{MS97}
U.-G.~Mei\ss ner and S.~Steininger  
{\it Nucl. Phys.}  {\bf B499}, 349 (1997).

\bibitem{Beane:2003xv}
S.R.~Beane and M.J.~Savage,
{\it Phys. Rev.} {\bf D68}, 114502 (2003).

\bibitem{Elizalde:1997jv}
E.~Elizalde,
{\it Commun. Math. Phys.} {\bf 198}, 83 (1998).

\bibitem{savagenotes}
M.J.~Savage, {\it unpublished notes}.

\end{thebibliography}
\end{document}